\documentclass[prb,twocolumn,showkeys,showpacs,preprintnumbers,amsmath,amssymb,floatfix]{revtex4}

\usepackage{graphicx}
\usepackage{dcolumn}
\usepackage{bm}

\begin{document}

% \preprint{PRB 2005}

\title{Effect of annealing on the depth profile of hole concentration in (Ga,Mn)As}

\author{W. Limmer}
 \email{wolfgang.limmer@physik.uni-ulm.de}
 \homepage{http://hlpsrv.physik.uni-ulm.de}
\author{A. Koeder}
\author{S. Frank}
\author{V. Avrutin}
\author{W. Schoch}
\author{R. Sauer}
\affiliation{Abteilung Halbleiterphysik, Universit\"at Ulm, D-89069 Ulm, Germany}
\author{K. Zuern}
\author{J. Eisenmenger}
\author{P. Ziemann}
\affiliation{Abteilung Festk\"orperphysik, Universit\"at Ulm, D-89069 Ulm, Germany}
\author{E. Peiner}
\author{A. Waag}
\affiliation{Institut f{\"u}r Halbleitertechnik, Technische Universit{\"a}t Braunschweig,
D-38023 Braunschweig, Germany}
% \date{\today}

\begin{abstract}
The effect of annealing at 250$^\circ$C on the carrier depth profile, Mn distribution, electrical
conductivity, and Curie temperature of (Ga,Mn)As layers with thicknesses $\geq$ 200 nm, grown by
molecular-beam epitaxy at low temperatures, is studied by a variety of analytical methods. The
vertical gradient in hole concentration, revealed by electrochemical capacitance-voltage profiling,
is shown to play a key role in the understanding of conductivity and magnetization data. The
gradient, basically already present in as-grown samples, is strongly influenced by post-growth
annealing. From secondary ion mass spectroscopy it can be concluded that, at least in thick layers,
the change in carrier depth profile and thus in conductivity is not primarily due to out-diffusion of
Mn interstitials during annealing. Two alternative possible models are discussed.
\end{abstract}



\maketitle

\section{Introduction}

Semiconductor-based spintronic technology, where both the electrical charge and the spin of carriers
are utilized for signal processing and storage, calls for the development of new ferromagnetic
materials. The III-V dilute magnetic semiconductor (Ga,Mn)As, being compatible with conventional
semiconductor technology, is considered a potential candidate for spintronics and has been intensely
studied during the past few years.\cite{Ohn98,Esc97,Mat98} The ferromagnetic properties of (Ga,Mn)As,
successfully explained within the Zener mean-field model\cite{Die01}, arise from the $S=5/2$ spins of
Mn atoms incorporated on Ga lattice sites. The ferromagnetic Mn-Mn coupling is mediated by delocalized
or weakly localized holes which are supplied by the Mn atoms acting as acceptors on Ga lattice sites.
The Mn spin system undergoes a ferromagnetic phase transition at the Curie temperature $T_{\mbox{c}}$
which is suggested to strongly depend on both, the Mn content and the hole concentration.\cite{Die01}
(Ga,Mn)As is grown by molecular-beam epitaxy (MBE) at low temperatures ($\sim$250$^\circ$C).
In most cases, $T_{\mbox{c}}$ is increased by post-growth annealing at temperatures near or even below
the growth temperature, resulting in values of up to 160 K in (Ga,Mn)As single layers, as reported so
far.\cite{Ku03,Pot01,Edm02,Soe03,Lim04,Edm04a,Chi03,Sto03} This is commonly explained by the removal
or rearrangement of highly unstable compensating defects such as Mn atoms on interstitial lattice
sites.\cite{Edm04a,Chi03,Sto03,Yu02,Bli03} The enhancement of the ferromagnetism associated with
low-temperature (LT) annealing is suppressed in the presence of a thin GaAs capping layer,
indicating that diffusion of defects towards the surface plays a crucial role in the annealing
process.\cite{Chi03,Sto03} It has been revealed by electrochemical capacitance-voltage (ECV)
profiling and Raman spectroscopy that in general the hole concentration $p$ is not constant
throughout the (Ga,Mn)As layer but exhibits a vertical gradient.\cite{Koe03} Starting at the
GaAs/(Ga,Mn)As interface, $p$ monotonously increases and reaches its maximum value near the sample
surface. This finding is in agreement with the results of spin-wave resonance experiments which
manifest the existence of a gradient in the magnetic properties.\cite{Goe03,Rap03}

For (Ga,Mn)As, it is well known that, within a limited temperature range, a slight increase of the
growth temperature results in a pronounced enhancement of both, $T_{\mbox{c}}$ and $p$.\cite{Shi99}
In addition, a significant increase of the surface temperature, induced by heat radiation from the
effusion cells, has been detected during LT MBE growth of GaAs.\cite{Tho97,Zha02} Therefore, we
suppose that besides annealing effects during growth a gradual increase of the surface temperature
due do free carrier absorption is likely to account for the observed gradient. Even though
disregarded by most other groups so far, the presence of the gradient seems to be a general
phenomenon, whereas its specific profile depends on several parameters such as sample thickness and
growth conditions. From the $p$-dependence of $T_{\mbox{c}}$ it becomes obvious that a pronounced
variation in $p$ along the growth direction must have a strong impact on the electrical and magnetic
properties of (GaMn)As.

In this work, the effect of annealing at 250$^\circ$C on the depth profile of hole
concentration in (Ga,Mn)As epilayers with thicknesses between 0.2 and 1.2 $\mu$m is
studied by ECV profiling, micro-Raman spectroscopy, conductivity measurements,
superconducting quantum interference device (SQUID) magnetization measurements, and
secondary ion mass spectroscopy (SIMS). It is shown that the gradient in the carrier
density plays a key role in the understanding of annealing-induced effects, such as the
increase in conductivity and Curie temperature. From our SIMS and ECV data we infer that,
at least in thick layers ($\sim$~1~$\mu$m), out-diffusion of Mn interstitials
is not the dominant mechanism for the annealing-induced
enhancement of the hole density, as recently proposed for thin layers ($\leq$
100 nm).\cite{Edm04a,Sto03} While most reports in the
literature are on (Ga,Mn)As epilayers with thicknesses in the range of 10-200 nm, where
the gradient is probably less pronounced, thicker samples, as considered in this work,
allow to map the depth profiles, e.g., of the hole and Mn concentration over larger
distances, and therefore to get more detailed information to the understanding of the
defect dynamics. Provided that the fundamental diffusion processes during annealing are
primarily independent of the layer thickness, the findings reported in this paper may be
helpful in the interpretation of experimental data obtained from much thinner samples as
well.

\section{Experimental details}

(Ga,Mn)As layers were grown in a RIBER 32 MBE machine on In-mounted semi-insulating VGF
GaAs(001) substrates using a conventional Knudsen cell and a hot-lip effusion cell to
provide the Ga and Mn fluxes, respectively. A valved arsenic cracker cell was used in the
non-cracking mode to supply As$_4$ with a maximum V/III flux ratio of about
3. First, a GaAs buffer layer around 100 nm thick was grown at a temperature of $T_s$ =
585$^\circ$C (conventional substrate temperature for GaAs), then the growth was
interrupted and $T_s$ was lowered to $\sim$250$^\circ$C. The Mn concentrations in the
$0.2-1.2$ $\mu$m-thick (Ga,Mn)As layers under study were determined by flux measurements,
which have been checked by elastic recoil detection measurements (ERD). For details about
ERD, see Ref.~\onlinecite{Dol98}.

ECV analyses were performed using a Bio-Rad PN4200 profiler. The electrolyte (250 ml aqueous solution
of 2.0 g NaOH + 9.3 g EDTA) is in contact with the semiconductor forming an electrolyte-semiconductor
diode. A low-resistance Ohmic contact can be established to the (Ga,Mn)As sample without metallization.
The potential of the sample is measured potentiometrically with reference to a saturated calomel
electrode. The admittance $Y$ of the electrolyte-semiconductor contact is determined by ac
measurements under reverse bias at $\omega_c$. Additionally, the bias voltage $V_m$ is wobbled at a
considerably lower modulation frequency to yield the differential admittance $dY/dV$. $V_m$ and
$\omega_c$ are selected to obtain optimum Schottky characteristics of the electrolyte-semiconductor
contact. The corresponding equivalent circuit which is implemented in the analysis software consists
of the space-charge layer capacitance $C$ and of resistances $R_p$ and $R_s$ connected in parallel
and in series, respectively. $R_s$ is obtained from the admittance data measured at two different
frequencies $\omega_c$. Using $Y$, $dY/dV$, and $R_s$, we can determine $C$ and $dC/dV$ yielding the
carrier concentration $N(w_d)$ at the edge of the depletion region of width $w_d$ according to
\begin{equation}
N(w_d) = -\frac{C^3}{e\varepsilon A^2}\left(\frac{dC}{dV} \right)^{-1}
\label{eq:Nwd}
\end{equation}
with
\begin{equation}
w_d = \frac{\varepsilon A}{C}\;,
\label{eq:wd}
\end{equation}
where $\varepsilon$ (0.12 nF/m) and $e$ are the permittivity of GaAs and the electron charge,
respectively. Controlled amounts of (Ga,Mn)As are removed in increments of few tens of nanometers by
passing a dc current $I$ between the anodically polarized semiconductor and a carbon
counter electrode. The removed layer thickness $w_e$ is calculated from the accumulated transferred
charge using Faraday's law of electrolysis:
\begin{equation}
w_e = \frac{M}{zF\rho A}\int Idt\;,
\label{eq:we}
\end{equation}
where $M$ (144.6), $z$ (6), $\rho$ (5.36 g/cm$^{3}$), and $F$ (96490 Asmol$^{-1}$) denote
the molecular weight, effective dissolution valence, and density of GaAs, and Faraday's constant,
respectively. Additionally, $w_e$ is controlled by mechanical surface tracing. The diode area $A$ of
0.005 to 0.008 cm$^2$ is defined by a plastic sealing ring. Normally, $A$ is not accurately known at
the beginning of an ECV profiling run, but is routinely measured subsequently. Relevant
recalculation procedures considering the measured values of $A$ and the series resistance $R_s$ are
implemented in the original Bio-Rad software of the PN 4200 ECV profiler. The measured ECV profiles
are reproducible within an uncertainty in the absolute values of about 15$\%$. More information
about ECV profiling can be found in Ref.~\onlinecite{Blo86}.

Hole concentrations in (Ga,Mn)As can also be estimated from Raman scattering by coupled plasmon-LO-phonon
modes.\cite{Lim02,Seo02} Therefore, micro-Raman measurements were performed at room temperature (RT) using
the 514-nm line of an Ar$^{+}$ laser as an excitation source. The Raman signals were detected in the
backscattering configuration $\bar z(x,y)z$ using a DILOR XY 800-mm triple-grating spectrometer with a
confocal entrance optics and a LN$_{2}$-cooled charge-coupled device detector. Further experimental
details of the micro-Raman measurements are given in Ref.~\onlinecite{Lim02}.

For the electrical measurements, Hall bars with Ti-AuPt-Au contacts were prepared on several pieces of
the cleaved samples without annealing. The contacts were checked to be Ohmic with negligibly low
resistance. The samples were annealed in air using a LINKAM THMS 600 heating chamber equipped with an
electrical feed through, which enabled us to perform {\it in situ} measurements of the conductivity.
They were mounted on a silver block which could be heated electrically or cooled by liquid nitrogen
over the temperature range from -200 to 300$^\circ$C within 2 minutes.

To measure depth profiles of the Mn fraction, SIMS experiments were performed using a
commercial Cameca ims4f-E6 spectrometer with Cs$^+$ as primary ion beam (net impact
energy 5.5 keV) at a sputter rate of about 1 nm/s. In the absence of appropriate
calibration standards for Mn, the quantitative analysis of the Mn fraction refers to the
flux measurement of sample B313, which gave us a value of 6$\%$. A detailed introduction
into the SIMS measuring method is given in Ref.~\onlinecite{Fuc90}.

The magnetization measurements were carried out in a QUANTUM DESIGN MPMS 5 SQUID magnetometer applying
an in-plane magnetic field of 5 mT.

\section{Results and discussion}

The phenomena discussed in this paper have been qualitatively observed in all
(Ga,Mn)As layers grown at V/III flux ratios $\leq$ 3, exhibiting thicknesses
between 200 nm and 1.2 $\mu$m. The specific influence of the V/III flux ratio on the
structural, electric, and magnetic properties of (Ga,Mn)As is not subject of this
work and will be discussed elsewhere. In the following we present experimental
results obtained from several pieces of a 1.2-$\mu$m-thick (Ga,Mn)As epilayer with
a Mn fraction of 6$\%$ (sample B313) and of a 240-nm-thick epilayer with a
Mn fraction of 4.5$\%$ (sample B352). These epilayers are representative for all
other samples investigated so far.

\begin{figure}[!]
\includegraphics[scale=0.5]{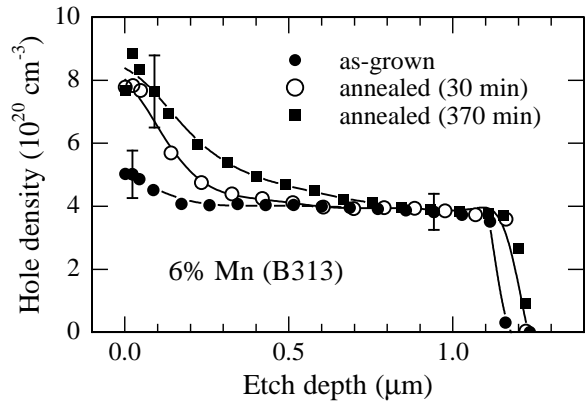}
\caption{\label{fig:ecv_B313} ECV profiles recorded from sample B313 before and after
annealing at 250$^\circ$C for 30 and 370 min.}
\end{figure}

As mentioned above, ECV profiling reveals the presence of a vertical gradient in the hole
concentration. Figure~\ref{fig:ecv_B313} depicts the ECV profiles of sample B313 before and after
annealing at 250$^\circ$C for 30 and 370 min. It is clearly seen that the gradient, already present
in the as-grown sample, is strongly enhanced by post-growth annealing, leading to a hole density
near the sample surface which is almost twice as high as near the GaAs/(Ga,Mn)As interface. As a
consequence, the electric and magnetic properties of the (Ga,Mn)As layer are affected in a dramatic
way, as will be shown below. Whereas the total hole concentration, averaged over the layer
thickness, increases from $(3.8 \pm 0.6) \times 10^{20}$ cm$^{-3}$ in the as-grown sample to $(4.3
\pm 0.6) \times 10^{20}$ cm$^{-3}$ in the 30-min-annealed sample and to $(4.9 \pm 0.7) \times
10^{20}$ cm$^{-3}$ in the 370-min-annealed sample, the local hole density near the sample surface
almost saturates after annealing for 30 min. The particular evolution of the carrier depth profile
in the course of the annealing seems to corroborate the assumption that out-diffusion of
compensating defects towards the surface accounts for the increase in hole density during
post-growth annealing.\cite{Edm04a,Chi03,Sto03}

\begin{figure}[!]
\includegraphics[scale=0.5]{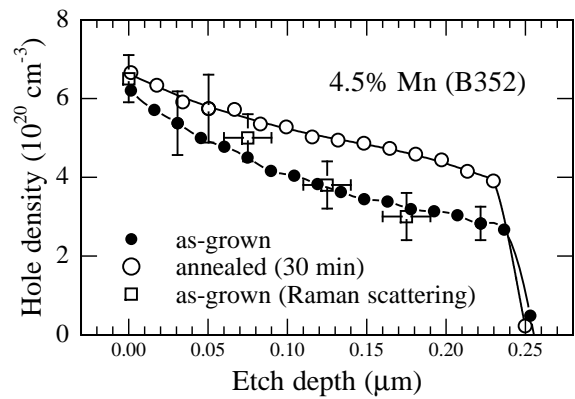}
\caption{\label{fig:ecv_B352} ECV profiles recorded from sample B352 before and after
annealing at 250$^\circ$C for 30 min. Depth-dependent hole concentrations determined from
Raman spectroscopy are depicted for comparison.}
\end{figure}

The ECV profiles of sample B352, recorded before and after annealing at 250$^\circ$C
for 30 min, are shown in Fig.~\ref{fig:ecv_B352}. The gradient in the as-grown sample is
much more pronounced than in B313 and flattens upon annealing with a concomitant increase in
the total hole concentration. The ECV profiles resemble those of sample B313 in the range
from the surface down to $\sim$~240 nm after annealing for 30 and 370 min.

In order to make sure that ECV profiles reflect the correct depth distribution of the hole
concentration, confirmative optical experiments were performed. To this end, steplike surface
profiles with flat terraces at different depths below the initial surface were prepared on several
(Ga,Mn)As layers by wet chemical etching. Micro-Raman spectra were taken from each of the individual
terraces and the carrier densities, obtained from a line-shape analysis of the Raman signals, were
compared with the ECV data. Figure~\ref{fig:ecv_B352} shows, as an example, such a comparison
for the as-grown sample B352. Within the error margin of the two analysis methods, the measured
depth profiles coincide almost perfectly.

\begin{figure}[!]
\includegraphics[scale=0.5]{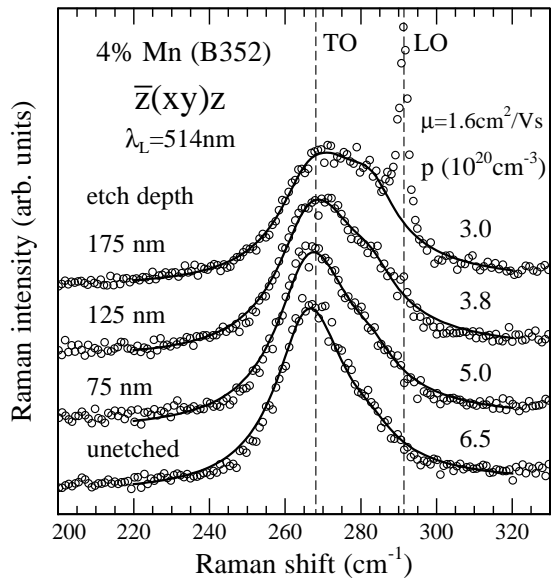}
\caption{\label{fig:raman_B352} Raman spectra recorded from sample B352 for different etch depths.}
\end{figure}

The Raman spectra corresponding to the data points in Fig.~\ref{fig:ecv_B352} are depicted in
Fig.~\ref{fig:raman_B352}. The high hole concentration in (Ga,Mn)As leads to the formation of a
phonon-like coupled mode of the longitudinal optical (LO) phonon and the overdamped hole
plasmon.\cite{Irm97a} With increasing hole concentration, this mode shifts from the frequency of the
LO phonon to that of the transverse optical (TO) phonon. It is clearly seen in Fig.~\ref{fig:raman_B352}
that the coupled mode broadens and shifts to higher frequencies with increasing etch depth, indicating
a decrease in the hole concentration. At an etch depth of 175 nm, the remaining layer thickness
nearly matches the information depth of $1/2\alpha \approx 50$ nm of the Raman measurement, where
$\alpha$ denotes the absorption coefficient at 514 nm wavelength of the Ar$^{+}$ laser. Therefore,
the narrow Raman line of the pure LO-phonon mode in the undoped substrate appears at 292 cm$^{-1}$.
The values for the hole densities were obtained from line-shape analyses of the Raman spectra using
a value for the hole mobility of 1.6 cm$^2$/Vs. The calculated line shapes are drawn as solid lines
in Fig.~\ref{fig:raman_B352}. Details concerning the calculation of Raman line shapes in heavily
p-doped semiconductors can be found in Refs.~\onlinecite{Lim02} and \onlinecite{Irm97a}.

\begin{figure}[!]
\includegraphics[scale=0.5]{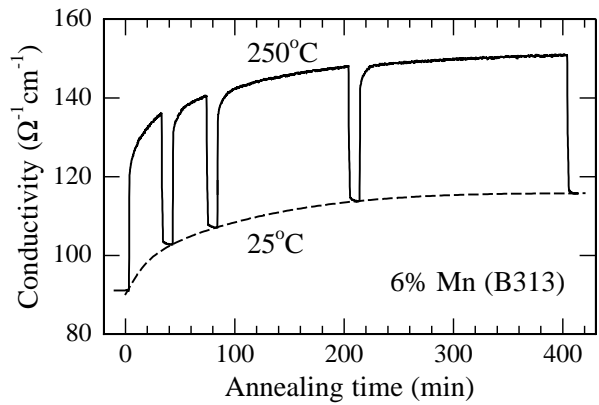}
\caption{\label{fig:sigma_lt_B313} {\it In situ} monitoring of the conductivity of sample B313 during
annealing at 250$^\circ$C.}
\end{figure}

Under the assumption that the hole mobility does not significantly change during the annealing process,
an increase in the hole concentration, as seen from Figs.~\ref{fig:ecv_B313} and \ref{fig:ecv_B352},
should result in an increase in the electrical conductivity of the (Ga,Mn)As layer. In fact, such an
increase has been observed by several authors.\cite{Pot01,Edm02,Edm04a,Lim04} In the present work,
the effect of annealing at 250$^\circ$C on the conductivity of sample B313 is shown in
Fig.~\ref{fig:sigma_lt_B313}.

Monitoring the conductivity by an {\it in situ} measurement for 400 min reveals that
the conductivity monotonously rises with a rate that gradually decreases with increasing annealing time.
After total annealing times of 30, 60, and 180 min, the annealing process was interrupted for 10 min
and the temperature was rapidly lowered to 25$^\circ$C in order to probe the RT conductivity (dashed line).
During the first 30 min, the RT conductivity increases from 91 to 103 $\Omega^{-1}$cm$^{-1}$. After
annealing for 370 min, the RT conductivity has almost saturated at a value of 116 $\Omega^{-1}$cm$^{-1}$.
According to these values, the conductivity is enhanced by factors of 1.13 and 1.27 upon annealing for
30 and 370 min, respectively. In contrast to Ref.~\onlinecite{Pot01}, no decrease of the conductivity
for annealing times longer than 2 h is observed. Note that the results of our {\it in situ}
measurements are in qualitative agreement with those obtained from much thinner (Ga,Mn)As
epilayers (10-100 nm) at lower annealing temperatures ($\leq$ 200$^\circ$C).\cite{Edm02,Edm04a}

We may now compare the annealing-induced increase of the conductivity with that in hole concentration.
From the averaged hole concentrations, derived above from the ECV profiles in Fig.~\ref{fig:ecv_B313},
we obtain an increase in hole density by factors of $1.13\pm 0.3$ and $1.29\pm 0.4$, which are in
excellent agreement with the values obtained for the conductivity. Thus, the assumption of at most a
small change in hole mobility during annealing is clearly confirmed. From the values of the conductivity
and the averaged carrier concentrations, effective hole mobilities of $1.5 \pm 0.3$ cm$^2$/Vs are deduced
for the as-grown as well as for the annealed samples.

\begin{figure}[!]
\includegraphics[scale=0.5]{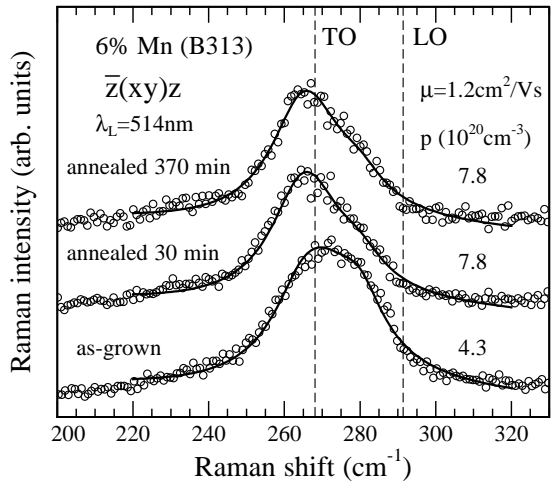}
\caption{\label{fig:raman_B313} Raman spectra recorded from sample B313 before and after
annealing at 250$^\circ$C. The solid lines are calculated line shapes.}
\end{figure}

Raman spectra, recorded from sample B313 before and after annealing at 250$^\circ$C for 30 and 370 min,
are shown in Fig.~\ref{fig:raman_B313}. Whereas a strong increase in the hole concentration within
the first 30 min of annealing can be deduced, the Raman spectrum recorded from the 370-min-annealed
sample does not significantly differ from that of the 30-min-annealed sample. This is in agreement
with the ECV profiles in Fig.~\ref{fig:ecv_B313}, which reveal that near the sample surface the hole
density is only slightly enhanced by annealing for more than 30 min. For the interpretation of the
Raman spectra one should keep in mind that the Raman signal, as already mentioned above, stems from
the near-surface region. The solid lines represent model calculations of the Raman line shapes using
a hole mobility of $\mu$ = 1.2 cm$^2$/Vs and hole concentrations of $4.3\times 10^{20}$ cm$^{-3}$
and $7.8\times 10^{20}$ cm$^{-3}$ for the as-grown and the annealed samples, respectively.

\begin{figure}[!]
\includegraphics[scale=0.5]{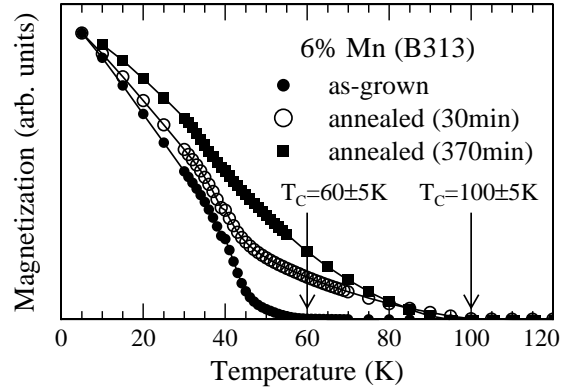}
\caption{\label{fig:squid_B313} Normalized magnetization of sample B313 as a function of
temperature before and after annealing.}
\end{figure}

According to the relation\cite{Die01}
\begin{equation}
T_{\mbox{c}}\propto x \times p^{1/3}
\label{eq:Tc}
\end{equation}
between Curie temperature $T_{\mbox{c}}$ and hole density $p$, with $x$ denoting the
concentration of magnetically active Mn ions on Ga sites, an increase in the hole density
should result in an enhancement of the Curie temperature. In fact, such an enhancement is
shown by the SQUID magnetization curves in Fig.~\ref{fig:squid_B313}, recorded from
sample B313 before and after annealing at 250$^\circ$C for 30 and 370 min.
Note that the curves are normalized to the values at 5 K, and thus, cannot be used to
obtain information about the influence of post-growth annealing on the saturation
magnetization. We suggest that the extended tails of the magnetization curves
arise from the vertical gradient of the hole density $p$ in the (Ga,Mn)As layer. According
to Eq.~(\ref{eq:Tc}), this gradient results in a depth-dependent Curie temperature
$T_{\mbox{c}}$, and thus the curves in Fig.~\ref{fig:squid_B313} can be viewed as
superpositions of individual magnetization curves. Then, the values of $T_{\mbox{c}}$
indicated by arrows have to be attributed to the near-surface region, similar to
the hole densities obtained from the Raman measurements. Whereas a Curie temperature
$T_{\mbox{c}}$ of $60 \pm 5$ K is deduced for the as-grown sample, a constant value of
$100 \pm 5$ K is obtained for the two annealed samples, yielding an increase of
$T_{\mbox{c}}$ by a factor of $1.7 \pm 0.3$. In contrast, an enhancement of
$T_{\mbox{c}}$ by a factor of only $1.2 \pm 0.1$ would have been expected from
Eq.~(\ref{eq:Tc}) and the ECV data measured near the sample surface, yielding an increase
of the local hole density by a factor of $1.7 \pm 0.5$. The discrepancy between the two
values may be explained by a reduction of the number of antiferromagnetically ordered Mn
atoms\cite{Bli03} during annealing.

\begin{figure}[!]
\includegraphics[scale=0.5]{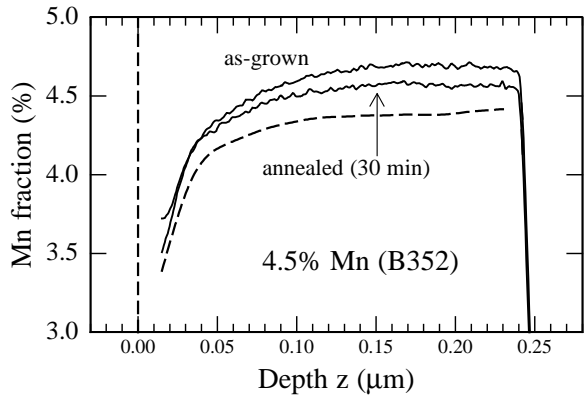}
\caption{\label{fig:sims_B352} Mn depth profiles of sample B352 before and after annealing
at 250$^\circ$C for 30 min measured by SIMS. The dashed line
represents the calculated Mn profile of the annealed sample using Eqs.~(\ref{eq:mnann})
and (\ref{eq:mnsur}).}
\end{figure}

\begin{figure}[!]
\includegraphics[scale=0.5]{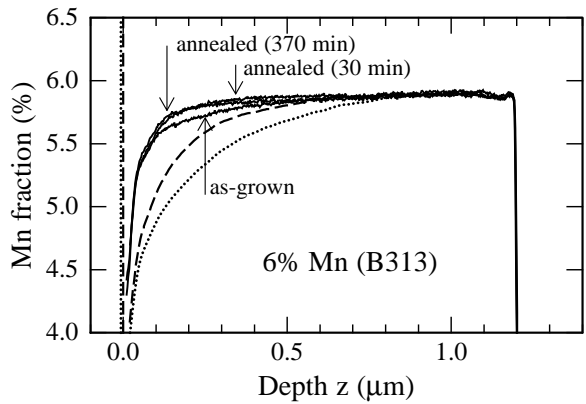}
\caption{\label{fig:sims_B313} Mn depth profiles of sample B313 before and after annealing
at 250$^\circ$C for 30 and 370 min measured by SIMS. The dashed and dotted lines
represent the calculated Mn profiles of the 30-min and 370-min-annealed sample, respectively,
using Eqs.~(\ref{eq:mnann}) and (\ref{eq:mnsur}).}
\end{figure}

The increase in hole density, conductivity, and Curie temperature upon post-growth
annealing is commonly explained by the removal of compensating defects in (Ga,Mn)As. The
results of ion channeling experiments point to a reduction of Mn interstitials (Mn$_I$),
acting as compensating double donors.\cite{Yu02} Based on the observation of Mn
accumulation at the sample surface and on theoretical calculations, this reduction has been
traced back to an out-diffusion of Mn$_I$ towards the surface followed by
oxidation.\cite{Edm04a,Sto03,Edm04b} In order to verify the latter suggestion for the
thick (Ga,Mn)As layers under study, the Mn depth profiles of the as-grown and the annealed
samples were experimentally determined by SIMS measurements recorded from the same sample
pieces as used for ECV profiling. The Mn profiles are depicted in Figs.~\ref{fig:sims_B352} and
\ref{fig:sims_B313} for B352 and B313, respectively. Whereas
Fig.~\ref{fig:sims_B352} suggests a slight annealing-induced lowering of the Mn fraction
in the 240-nm-thick sample, no significant difference between the three profiles of the
1.2-$\mu$m-thick sample can be identified in Fig.~\ref{fig:sims_B313}. Inevitably the
question arises if an out-diffusion of Mn$_I$, necessary to completely account for the
observed ECV profiles, would result in a significant and measureable change in the Mn
depth profile at all. Therefore, simple considerations are made in the following to estimate
the annealing-induced changes in the Mn depth profiles expected for the extreme case that the
increase in the hole densities in Figs.~\ref{fig:ecv_B313} and \ref{fig:ecv_B352} was entirely
due to the out-diffusion of Mn$_I$.

The hole concentration $p$ is given by the density $[Mn_{Ga}]$ of substitutional Mn acceptors
(Mn$_{Ga}$)
minus the total density of compensating donors:
\begin{equation}
p = [Mn_{Ga}] - C \times [Mn_I] - d\;,
\label{eq:p}
\end{equation}
where $[Mn_I]$ denotes the density of Mn$_I$ and $d$ the density of all other compensating
defects. Note that the terms in Eq.~(\ref{eq:p}) are local quantities which may in general
vary with depth within the sample. The factor $C$ accounts for the fraction of Mn$_I$ acting
as donors as well as for the corresponding valency (2 for double donors). Therefore, the
inequality $C\leq 2$ holds, where the sign of equality applies for the case that all Mn$_I$
act as double donors. The total density $[Mn]$ of Mn atoms is given by
\begin{equation}
[Mn] = [Mn_{Ga}] + [Mn_I] + [Mn_{ia}]\;,
\label{eq:mn}
\end{equation}
with $[Mn_{ia}]$ denoting the density of electrically inactive Mn atoms. Let us now
consider the hypothetical case that the increase in $p$, revealed by ECV profiling, is solely
due to out-diffusion of Mn$_I$ and that inside the (Ga,Mn)As layer $d$,
$[Mn_{Ga}]$, and $[Mn_{ia}]$ remain unaffected upon annealing. Then, using Eqs.~(\ref{eq:p})
and (\ref{eq:mn}), the local density $[Mn]_{ann}$ of Mn atoms inside the annealed sample
$(z > 0)$ is given by
\begin{equation}
[Mn]_{ann} = [Mn]_{ag} - (p_{ann}-p_{ag})/C\;,
\label{eq:mnann}
\end{equation}
where the subscripts $ann$ and $ag$ stand for annealed and as-grown, respectively. The
increase in the Mn concentration on the surface ($z$=0) due to the accumulation of out-diffused
Mn$_I$ can be calculated from the conservation of the total amount of Mn as follows:
\begin{equation}
\left( [Mn]_{ann}-[Mn]_{ag} \right)_{z=0} \times \delta = \int_0^t dz (p_{ann}-p_{ag})/C\;,
\label{eq:mnsur}
\end{equation}
where $t$ denotes the thickness of the as-grown (Ga,Mn)As epilayer and $\delta$ the
thickness of the Mn surface layer, which is of the order of nm.

The dashed and dotted lines in Figs.~\ref{fig:sims_B352} and \ref{fig:sims_B313}
represent the calculated Mn profiles for the 30-min and 370-min-annealed samples, respectively,
using Eqs.~(\ref{eq:mnann}) and (\ref{eq:mnsur}) with $C=2$, and taking into account that a Mn
fraction of 1$\%$ corresponds to a concentration of Mn atoms of $2.2\times 10^{20}$ cm$^{-3}$.
For $[Mn]_{ag}$, the experimental SIMS profiles of the as-grown samples were used, while $p_{ag}$
and $p_{ann}$ were taken from Figs.~\ref{fig:ecv_B313} and \ref{fig:ecv_B352}. In the case
of sample B352 the calculated $[Mn]_{ann}$ curve inside the (Ga,Mn)As layer has qualitatively
the same form as the measured Mn profile, but quantitatively the annealing-induced decrease in
Mn fraction is about twice as strong. Note that our estimate would yield an even more pronounced
reduction of the Mn fraction if a value less than 2 had been used for the factor $C$. Thus, it
seems that in the 240-nm-thick sample only a small portion of the enhancement in hole density is
actually due to Mn$_I$ out-diffusion. In the case of sample B313 the calculated $[Mn]_{ann}$
curves inside the epilayer show a clear deviation from the Mn distribution in the as-grown sample,
beginning at a depth of about 0.5-0.7 $\mu$m and increasing strongly towards the surface.
This behavior is not even qualitatively reflected by the measured Mn profiles. Therefore,
we conclude that at least in thick ($\sim$~1~$\mu$m) samples the observed increase in hole
concentration in the bulk is not primarily due to out-diffusion of Mn$_I$. This finding is in
agreement with the low diffusivity of Mn$_I$ at 250$^\circ$C derived in Ref.~\onlinecite{Edm04a}.
Moreover, the calculated narrow peaks at the surfaces, arising from the accumulation of
out-diffused Mn$_I$, are not seen in the measured SIMS profiles. Calculation yields
$\left([Mn]_{ann}-[Mn]_{ag} \right)_{z=0} \geq 6\%$ for B352 and
$\left([Mn]_{ann}-[Mn]_{ag} \right)_{z=0} \geq 11\%$ for B313 with $\delta \leq 10$ nm.
Note however that the SIMS data presented in Fig.~\ref{fig:sims_B352} and \ref{fig:sims_B313}
may be too rough to resolve an extremely thin Mn-rich oxide layer on the surface, if present.

Our findings discussed above do not contradict the model deduced for much thinner (Ga,Mn)As
layers ($\lesssim$ 100 nm) that out-diffusion of Mn$_I$ plays a dominant role. In fact, the
experimental data available so far suggest that, due to the low diffusivity of Mn$_I$ at
250$^\circ$C, a significant out-diffusion of Mn$_I$ occurs in the near-surface region, whereas
in the bulk the highly unstable Mn$_I$ atoms rearrange and form electrically inactive randomly
distributed precipitates, as proposed in Ref.~\onlinecite{Yu02}. Within the framework of our simple
considerations above, this means that in the bulk the rise in hole concentration $p$ is mainly due
to a local increase of $[Mn_{ia}]$ and/or $[Mn_{Ga}]$ at the expense of $[Mn_I]$. From
the ECV profiles in Fig.~\ref{fig:ecv_B313} it follows that this rearrangement of $[Mn_I]$ does
not take place homogeneously throughout the whole (Ga,Mn)As layer, leading to an overall upshift
of the hole density, but exhibits a pronounced dependence on sample depth. At present, there is
no final explanation for this particular evolution of the carrier depth profile on annealing which
seems to be governed by a diffusion-based mechanism. Further experiments such as depth-resolved
investigations on the Mn$_I$ distribution and theoretical studies have to be performed in the
future to clarify this point. Yet, we finally close with a brief discussion of two potential
mechanisms.

The first mechanism is based on a weak depth-dependent diffusion of Mn$_I$ towards the surface
associated possibly with a change of lattice site location. Initiated by an out-diffusion of Mn$_I$
in the near-surface region, Mn$_I$ atoms deeper in the bulk successively diffuse towards the surface
while forming electrically inactive Mn clusters and/or MnAs. Due to the low diffusivity of Mn$_I$,
this process gradually becomes ineffective with increasing sample depth. Thus, depending on the
annealing time, different diffusion-like profiles of Mn$_I$ evolve which result in the hole density
profiles shown in Figs.~\ref{fig:ecv_B313} and \ref{fig:ecv_B352}. This explanation also accounts
for the fact that, compared to the total amount of Mn inside the (Ga,Mn)As layer, only a negligibly
small fraction of Mn effectively migrates out of the bulk.

The second mechanism is more hypothetical and rests upon an assumed out-diffusion of highly mobile,
possibly compensating, defects other than Mn$_I$, acting at least partially as a trigger for the
rearrangement of Mn$_I$. Arsenic located in interstitial positions (As$_I$) may be a potential
candidate for such a defect. Similar to LT GaAs\cite{Pri95,Liu95,Hag94} the LT MBE growth of
(Ga,Mn)As leads to the incorporation of excess As up to 2$\%$, acting at least partially as
compensating donor defects.\cite{Mat98,Yu02} Even though controversially discussed, several authors
suggest that in LT GaAs a considerable fraction of the excess As atoms is located in interstitial
positions, acting as highly mobile defects, while the rest of the excess As is believed to be in
antisite positions (As$_{Ga}$).\cite{Yu92,Hoz95,New93} As$_{Ga}$ defects are known to remain stable
up to 450$^\circ$C and are therefore not expected to play a significant role in the physical processes
taking place at 250$^\circ$C.\cite{Bli92} Whereas a rearrangement of Mn$_I$ could be revealed in
(Ga,Mn)As by combined channeling Rutherford backscattering and by particle-induced x-ray emission experiments\cite{Yu02}, out-diffusion of As is difficult to verify. As atoms, which diffuse from
interstitial sites to the sample surface during annealing, efficiently desorb at the surface and
are therefore hardly detected by surface sensitive methods.

\section{Summary}

The depth profile of the hole concentration in thick ($\geq$ 200 nm) MBE-grown (Ga,Mn)As layers,
measured by ECV profiling, as well as its strong change upon post-growth annealing at 250$^\circ$C has
been shown to play a key role in the interpretation of conductivity and magnetization data. The
annealing-induced increase in the total hole concentration, derived from ECV profiling, is in excellent quantitative agreement with the change in electrical conductivity measured {\it in situ} during annealing.
The pronounced enhancement of the hole density near the sample surface, confirmed by micro-Raman
measurements, is accompanied by a distinct increase of the Curie temperature. The particular evolution of
the measured ECV profiles under continued annealing suggests that diffusion processes play a major role
in the post-growth annealing of (Ga,Mn)As. From a comparison between the ECV profiles and the Mn
distributions determined by SIMS, it is concluded that, in contrast to thin layers, the increase
in hole density upon post-growth annealing in thick ($\sim$~1~$\mu$m) samples is not primarily due to out-diffusion of Mn$_I$. We suppose that a depth-dependent rearrangement of Mn$_I$ in the bulk, initiated
by an out-diffusion of Mn$_I$ from the near-surface region, accounts for the change in the
hole-density profile. Alternatively, a process based on the out-diffusion of other highly mobile
defects, such as As$_I$, has been tentatively discussed.

\begin{acknowledgments}
The authors acknowledge financial support by the Deutsche Forschungsgemeinschaft, DFG Wa 840/4.
\end{acknowledgments}

\end{document}